\documentclass[reqno,11pt,centertags]{amsart}
\usepackage{amssymb,upref,amsfonts,bm}

\usepackage{amsmath}
\usepackage{mathtools}
\usepackage{cite}
\allowdisplaybreaks \numberwithin{equation}{section}

\usepackage{graphicx,lastpage}
\begin{document}

\newcommand{\onlinecite}[1]{\hspace{-1 ex} \nocite{#1}\citenum{#1}}

\title[Virial Expansion of Ideal Quantum Gases in Arbitrary Dimensions]{Properties of the Virial Expansion
and Equation of State of Ideal Quantum Gases in Arbitrary Dimensions}

\author[K{\aa}re Olaussen]{K{\aa}re Olaussen}
\address{Institutt for fysikk,
Norges Teknisk-Naturvitenskapelige Universitet, 
N--7491 Trondheim, Norway}
\email{Kare.Olaussen@ntnu.no}

\author[Asle Sudb{\o}]{Asle Sudb{\o}}
\address{Institutt for fysikk,
Norges Teknisk-Naturvitenskapelige Universitet, 
N--7491 Trondheim, Norway}
\email{Asle.Sudbo@ntnu.no}

\dedicatory{Dedicated  to Johan S. H{\o}ye on the occation of his 70th birthday}
\thanks{}
\date{May 8, 2014: Published in \emph{Transactions of The Royal Norwegian Society
		of Sciences and Letters}, 2014(3) 115--135}

\begin{abstract}

The virial expansion of ideal quantum gases reveals some interesting and amusing properties when considered as a function of dimensionality $d$. 
In particular, the convergence radius $\rho_c(d)$ of the expansion is particulary large at
{\em exactly\/} $d=3$ dimensions,
\(
   \rho_c(3)
    = 7.1068\ldots \times \lim_{d\to3} \rho_c(d)
\).
The same phenomenon occurs in a few other special (non-integer) dimensions.
We explain the origin of these facts, and discuss more generally the
structure of singularities governing the asymptotic behavior of
the ideal gas virial expansion.

\end{abstract}

\maketitle



\section{Introduction}


To cite an authoritative source,
{\em the treatment of Bose-Einstein and
Fermi-Dirac perfect gases can
be made in an extremely simple manner\/}\cite{JEMayer1}.
The topic has been known since the discovery
of quantum statistics\cite{Bose,Einstein,EFermi,PDirac}.
One might think that from a theoretical perspective
there is absolutely nothing new to discover about it.
Nonetheless, in this paper we report some  rather
amusing behavior found when investigating how
the virial expansion and other properties
depend on dimensionality $d$ of the system.
Here, $d$ is an effective dimension given by $d = 2 D/\nu$, where $D$ is the dimension of 
physical space and $\nu$ is the power of wave number 
entering the dispersion relation of the excitations in question (see below). 
For nonrelativistic massive particles, we have $d=D$. For massless fermions (a reasonable
description of neutrinos and electrons in topological insulators), we have $d=6$, 
and for the asymptotic low-energy spectrum of electrons in graphene, 
we have $d=4$. Moreover Bose-Einstein condensates with effective spatial dimensions $D$
ranging from 0 to 3 are now routinely made. Such systems are of considerable current interest.
While interactions obviously play a role to varying degrees in the above mentioned systems,  this has nevertheless
motivated us to revisit the behavior of ideal quantum gases as a function of dimensionality with an emphasis on 
the analytical structure of the equations of state and other thermodynamical quantities. The analytical
structure reveals itself as surprisingly intricate and complex as dimension is varied.     

For any dimension $d$ the
equation of state of an ideal quantum gas has a nonzero radius of
convergence at $\rho=0$. It can therefore be
analytically continued to a complete
Riemann surface with much ($d$-dependent)
mathematically interesting structure.
We have investigated how and where singularities
occur on this Riemann surface, how they change
character and position with changing $d$, and
how that governs the asymptotic behavior of
virial coefficients. We have even found
cases where the analytically continued equation
of state reappears in a new form, meaningful from the
requirements that both density $\rho$ and pressure $p$ are real
for physical values of the chemical potential $\mu$ (but
usually pathological with regard to more detailed
physical behavior). To our knowledge this has not 
previously been reported in the literature.\cite{Statphys20}

An ideal Bose gas undergoes a condensation in which the
zero-momentum ground state becomes macroscopically occupied
above a critical density (or theoretically equivalent
below a critical temperature) provided the dimensionality $d >2$. 
This is a phenomenon accompanied by true non-analyticities 
in thermodynamic potentials (however without the appearance of collective modes). 
One would then expect on general grounds that any expansion of physical 
quantities (e.g. pressure) in powers of  density would have a convergence 
radius given by the critical density or smaller.

On the other hand, it has been  known for a long time that the
virial expansion of an ideal non-relativistic Bose gas in $d=3$ 
dimensions possesses the peculiar and somewhat surprising property of 
having a convergence radius $\rho_c(3)$ which is much larger than the 
critical density for Bose--Einstein condensation,
$\rho_{\text{BE}}(3)=\zeta(\frac{3}{2})\Lambda_T^{-3}$
(where $\Lambda_T$ is the thermal de Broglie wavelength, and $\zeta$ is 
the Riemann zeta function).

This was first demonstrated by Fuchs\cite{Fuchs},
following  a conjecture by Widom\cite{BWidom}.
Jensen and Hemmer\cite{Jenssen_Hemmer} (see also
Ziff and Kincaid\cite{Ziff}) estimated that
$\rho_c(3)\approx 7\times\rho_{\text{BE}}(3)$, based on numerical
evaluation of the first 116 virial coefficients. The computation
is by no means a trivial one --- the $n$'th virial coefficient
has a magnitude of order $10^{-1.27 n}$, the remainder of
cancellations between terms of order 1. A reliable evaluation
of the 116'th coefficient required computations to be carried out to 
160 digits accuracy --- a quite formidable task in 1971.
In the above works, the reason for the existence of such
a surprisingly large convergence radius was not discussed.

In $d=2$ dimensions the situation is the opposite, $\rho_{\text{BE}}(2)$ is 
infinite while $\rho_c(2)=2\pi\Lambda_T^{-2}$ (as can be seen from the repeatedly 
discovered exact virial expansion in two dimensions\cite{Ziff,Sen,Susanne,SusanneEtAl}). 
This led us to investigate the dependence on dimensionality in
more detail. This had previously
been done by Ziff and Kincaid\cite{Ziff} (they focused on
integer dimensions only, but some of their results are valid
for arbitrary dimensions). Actually, our parameter $d$ need
not correspond to the physical dimension of the system ---
what matters is that the density of states per energy
interval scales like
\begin{equation}
  g(\varepsilon) \sim \varepsilon^{{d}/{2}-1},
\end{equation}
and that it is possible to define pressure and density
as position independent quantities.
For excitations with dispersion relation
$\varepsilon(\bm{p}) \sim \vert\bm{p}\vert^\nu$
in $D$ physical dimensions we find ${d}={2 D}/{\nu}$.
Only for a standard nonrelativistic spectrum will $d$ correspond
to the physical dimension. With excitations living on a fractal
subset of physical space we may also obtain a non-integer $d$.
Although there may be many interesting physical realizations,
the purpose of varying $d$ continuously in this paper
is mainly to obtain a more coherent and connected picture
of the behavior of the virial expansion as dimensionality is varied.

The rest of this paper is organized as follows. In Section
\ref{Virial}  we first discuss a peculiar symmetry between
fermions and bosons which occur in the case of ideal quantum
gases, its origin, and how it extends to interacting systems.
Next we discuss some general aspects of the virial expansion,
and display the first few term of the ideal Bose gas expansion
for general dimension $d$. 

Section~\ref{NumericalExperiments} describes our numerical exploration
of the virial coefficients $A_n$ as function of dimension $d$. 
Each $A_n$ has a number of zeros. They reveal intriguing patterns
which inspired many conjectures (and eventually this whole research).

In Section~\ref{SpecialCases} we consider the
special cases of $d=0$ (quantum dot), $d=2$ (confined  layer), and
$d=3$. Apart from their physical applications, $d=0$ and $2$
are interesting since in those cases the relation between fugacity
$z$ and density $\rho$ can be inverted exactly. Hence, they allow for
a more explicit and detailed analysis, and provide boundary conditions
which must be obeyed by the general expansions. The physically most
important system, $d=3$, is peculiar in that the equation of state
(i.e., pressure $p$ as function of $\rho$) extends analytically beyond
the density $\rho_{\text{BE}}$ for Bose-Einstein condensation. This analytic behavior
also holds for other thermodynamic quantities like the chemical
potential (the density fluctuations exhibit a pole singularity at $\rho_{\text{BE}}$).
The behavior for $\rho > \rho_{\text{BE}}$ is easily seen to be unphysical, and 
is not related to the correct behavior of the condensed state. 

Section~\ref{RiemannSurface} is a mathematical discussion of how the parametric
representation~(\ref{ImplicitEquationOfState}) can be extended to the Riemann surface 
of the polylogarithmic functions involved, leading to the general parametric 
representation~(\ref{GeneralEoS}). The sheets of the complete Riemann surface are labeled 
by an infinite-dimensional vector $\bm{k}$ of integers. On an infinite subset of these 
sheets the parametric representation~(\ref{GeneralEoS}) provide a candidate equation of 
state, related to the usual (low-density) Bose equation of state by analytic continuation. 

Section~\ref{TravelAccount} is a first account of our travel on the
Riemann surface of the equation of state, first locating the
singularities governing the radius of convergence of the virial
expansion for $d=3$, and next exploring how these singularities move
as $d$ is changed.

In Section~\ref{AsymptoticCoefficients} we demonstrate 
how knowledge of the closest singularities of $p(\rho)$,
and the behavior of $p(\rho)$ around these, can be used to provide an
accurate analytical representation of the asymptotic behavior of
the virial coefficients. The analytic prediction compares very well
with numerically calculated coefficients.

In Section~\ref{NearBEC} the behavior near the Bose-Einstein condensation point 
is analysed, revealing why the singularity in the equation of state vanishes at 
this point for dimensions $d=2+2/(m+1)$ (with $m=1 \ldots 6$). 

Section~\ref{ExploringDimension} is a
second account of our exploration of the Riemann surface, with
focus on how singularities of the equation of state flow when
$d$ is changed, and in particular the behavior of this flow as
$d\to 2$ and $d\to 0$. The singularities are in general of square root
type. However, for $d=2$ the singularities are logarithmic; hence each
of them must be formed by an infinite number of coalescing square root
singularities as $d\to2$. Moreover, (infinitely) many pairs of square
root singularities flow towards $\rho=0$  as $d\to0$, where none
can be seen in the explicit equation of state. The annhilation
mechanism seems to be similar to the one which is operative in the disappearance of two square root
singularities in $\sqrt{z^2-\varepsilon^2}$ when $\varepsilon\to 0$.

We close the paper with a few remarks in Section~\ref{Conclusions}.

\section{Virial expansions for ideal quantum gases\label{Virial}}

Consider a $d$-dimensional volume $V=L^d$ filled with identical nonrelativistic, noninteracting 
mass-$m$ quantum mechanical particles. Without internal degrees of freedom, the number of single-particle 
states in a tiny interval $\text{d}\varepsilon$ around $\varepsilon$ is
\begin{align}
    \frac{V}{\Gamma(d/2)}\left( \frac{2\pi m\varepsilon}{h^2}\right)^{d/2}\,
  \frac{\text{d}\varepsilon}{\varepsilon} \equiv V\,g(\varepsilon)\,\text{d}\varepsilon
\label{dosd}
\end{align}
when $\varepsilon >0$ (otherwise zero). Here, the surface of a unit sphere in $d$ dimensions is set to 
$S_{d-1} = 2\,\pi^{d/2}/\Gamma(d/2)$ in general, and we have assumed that $V\to\infty$ in a regular manner. 
At a given temperature $T$ we choose to measure energy in units of $k_B T$, and length in units of the 
thermal wavelength $\Lambda_T = (2\pi mk_BT/h^2)^{-1/2}$. For bosons in the grand canonical ensemble 
the equation of state is given in implicit form by 
\begin{align}
  p &= \frac{1}{\Gamma(1+d/2)} \int_{0}^{\infty} \frac{\text{d}\varepsilon}{\varepsilon}
  \frac{z\,\varepsilon^{1+d/2}}{e^{\varepsilon} - z}, \nonumber \\[-1.6ex] 
  \phantom{\cdot}   \label{ImplicitEquationOfState}\\[-1.6ex]
  \rho &= \frac{1}{\Gamma(d/2)} \int_{0}^{\infty} \frac{\text{d}\varepsilon}{\varepsilon}  
  \frac{z\,\varepsilon^{d/2}}{e^{\varepsilon} - z},\nonumber
\end{align}
where $z=\text{e}^{\mu}$  is the fugacity and $\mu$ the chemical potential.
The expressions for fermions are the same with the replacements
\begin{eqnarray}
   \rho \rightarrow -\rho,\quad
   z    \rightarrow -z,\quad
   p    \rightarrow -p.
   \label{BoseFermiSymmetry}
\end{eqnarray}
It is intriguing that the {\em same\/} functions describe both bosons and
fermions, only in different parameter ranges.

However, the generalization
of equation~(\ref{BoseFermiSymmetry}) to interacting systems is less direct.
In a functional integral formalism, the fugacity relation between bosons and
fermions is related to the rule that  for bosons one should integrate
over classical fields $\varphi$
which are periodic in the imaginary time ($\tau$) direction, and for fermions over
Grassmann fields $\chi$ which are antiperiodic.
This difference can be eliminated by a $\tau$-dependent gauge transformation at the
cost of transforming $\mu \to \mu + (2n+1)\pi\text{i}$, equivalent
to $z\to -z$. Without interactions the relation between pressures $p$ is
a consequence of the relation between classical and Grassmann gaussian integrals,
\begin{align*}
    \text{e}^{pV} &= \int \mathcal{D}\varphi^{*} \mathcal{D}\varphi\; 
    \text{e}^{-\varphi^{*}A\varphi} \propto \text{det}^{-1} A,\\
    \text{e}^{pV} &= \int \mathcal{D}\chi^{*} \mathcal{D}\chi\;
    \text{e}^{-\chi^{*}A\chi} \propto \text{det}\, A,
\end{align*}
which in the interacting case generalizes to the diagrammatic rule
of one extra minus-sign per fermion loop. I.e., if we for a
model with interactions know the complete loop expansion of the bosonic
functional integral, with $p_L(z)$ being the $L$-loop contribution, 
then relation (\ref{BoseFermiSymmetry}) generalized to
\begin{equation}
    p_f(z) = \sum_L (-1)^{fL} p_{L}((-1)^f z),\quad
    \rho_f(z) = z\frac{\text{d}}{\text{d}z} p_f(z),
\end{equation}
where $f=0$ for bosons and $f=1$ for fermions.

In the following, we return to the noninteracting model and we will mainly consider bosons in 
explicit expressions. The integrals (\ref{ImplicitEquationOfState}) can be written as power series in $z$
(Mayer expansions),
\begin{align}
  p &= z+\sum_{\ell=2}^{\infty}
  \frac{z^\ell}{\ell^{1+d/2}}  \equiv z + \sum_{\ell=2}^\infty  b_{\ell} z^{\ell},\nonumber\\[-1.75ex]
  \phantom{\cdot}\label{MayerExpansion}\\[-1.75ex]
  \rho &= z+\sum_{\ell=2}^{\infty}\frac{z^\ell}{\ell^{d/2}}  = z + \sum_{\ell=2}^\infty  \ell b_{\ell} z^\ell.\nonumber
\end{align}
The fermion expressions are obtained by $b_\ell \to (-)^{\ell+1}\, b_\ell$. By eliminating $z$ and expressing $p$ in 
terms of $\rho$ we obtain the virial expansion,
\begin{equation}
  p = \rho + \sum_{n=2}^{\infty} A_n\, \rho^n,
\end{equation}
where $A_n$ are the virial coefficients. The fermion cofficients are related to the bosonic ones by 
$A_n \to (-1)^{n+1} A_n$. Therefore, the bosons and fermions have the same radius of convergence
for their virial expansions, governed by singularities which are related by inversion, $\rho \to -\rho$. 
Due to reality conditions both cases also have reflection symmetry under $\text{Im}\,\rho \to -\text{Im}\,\rho$. 
Thus, singularities outside the real axis will occur in complex conjugate pairs.

The virial coefficients can be calculated by order-by-order inversion of the Mayer series~(\ref{MayerExpansion}). 
They can also be found through an explicit algorithm\cite{JEMayer2}. For the latter, define the power series
\begin{equation}
   g(z) = \sum_{\ell=1}^\infty (\ell+1) b_{\ell+1} z^\ell,
\end{equation}
and coefficients $C_{nm}$ through the generating function
\begin{equation}
   \frac{t\,g(z)}{1+t\,g(z)} = -\sum_{n=1}^\infty \sum_{m=1}^n C_{nm} z^n t^m.
\end{equation}
An alternative way to describe $C_{nm}$ is as follows:
Define $v_n=-(n+1) b_{n+1}$. Then
\[
 C_{nm}={\sum}'
 \left(\stackrel{\displaystyle n}{\nu_1\,\nu_2\,\nu_3\cdots}\right)\,
 v_1^{\nu_1} v_2^{\nu_2} v_3^{\nu_3}\cdots
 \]
where the sum is over all sets $\{\nu_1\,\nu_2\,\nu_3\cdots\}$
of non-negative integers 
such that $\sum_k \nu_k=m$ and $\sum_k k\nu_k=n$.
In any case the virial coefficients are
\begin{equation}
   A_{n+1} = \frac{1}{n+1} \sum_{m=1}^n 
   \left(\begin{array}{c}n+m-1\\m\end{array}\right)\,C_{nm}.
  \label{MayerPressureExpansion}
\end{equation}
Likewise, the fugacity can be expressed in terms of density as
\begin{equation}
   z = \rho + \sum_{n=2}^{\infty} B_n\,\rho^n,
\end{equation}
where
\begin{equation}
   B_{n+1} = \frac{1}{n+1} \sum_{m=1}^n 
   \left(\begin{array}{c}n+m\\m\end{array}\right)\,C_{nm}.
  \label{MayerFugacityExpansion}
\end{equation}
%
\begin{figure}
  \includegraphics[]{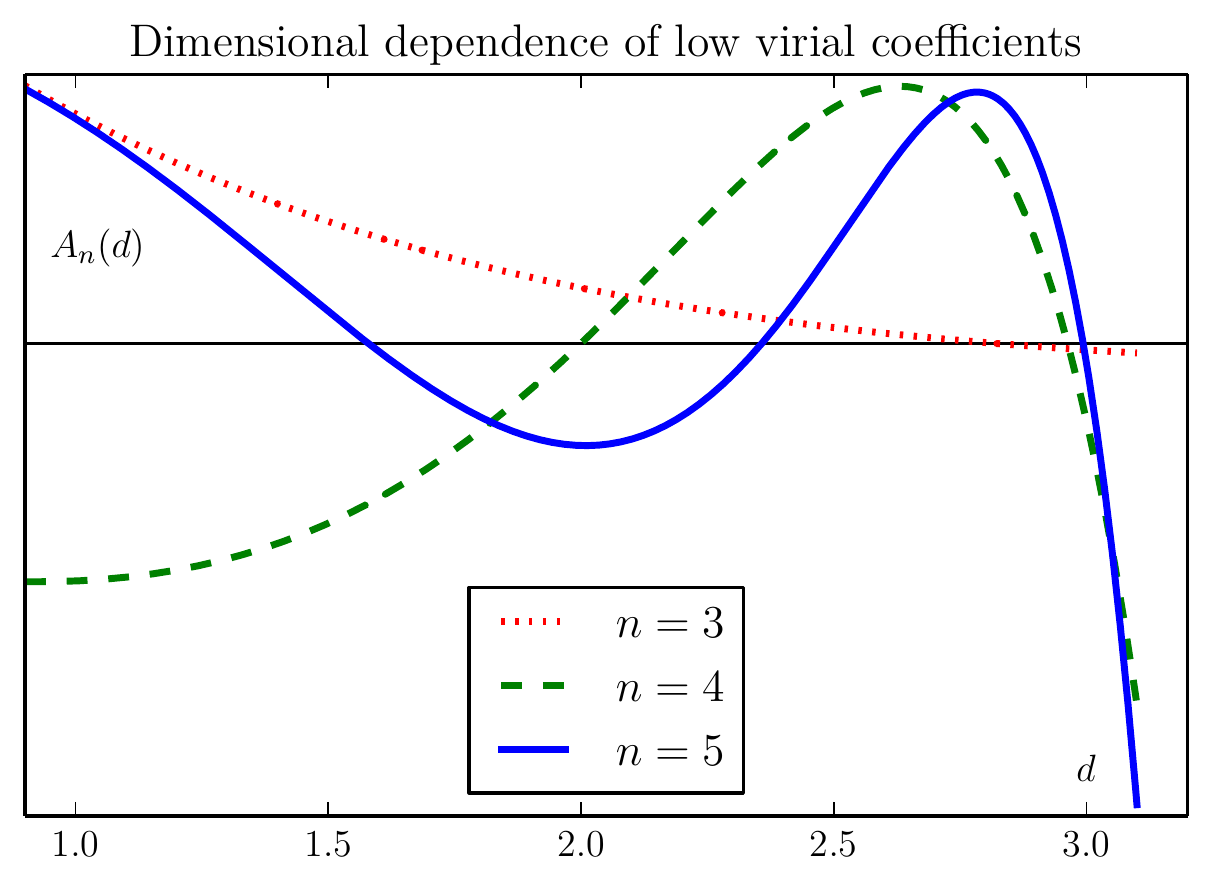}
  \caption{\label{figA3to5}
    The $3^{\text{rd}}$ to $5^{\text{th}}$ virial coefficients
  as function of dimension $d$. To adjust all curves to the same plot we have multiplied $A_n$ by
  expressions $\text{e}^{\alpha +\beta d}$, with real $n$-dependent coefficients
  $\alpha$ and $\beta$. Note that $A_n(d)$ has a zero which rapidly approaches
  $d=3$ as $n$ increases.} 
\end{figure}
%

The first virial coefficients are explicitly (with $x = 1+{d}/{2}$),
\begin{align}
   A_2(d) &= -2^{-x},\nonumber\\
   A_3(d) &= 4 \cdot 4^{-x} -2 \cdot 3^{-x},\nonumber\\
   A_4(d) &= -20 \cdot 8^{-x} +18 \cdot 6^{-x} -3 \cdot 4^{-x},\label{ExplicitCoefficients}\\
   A_5(d) &= 112 \cdot 16^{-x} -144 \cdot 12^{-x} +\nonumber\\
   &\phantom{=-} 18 \cdot 9^{-x} +32 \cdot 8^{-x} -4 \cdot 5^{-x}.\nonumber
\end{align}
They are plotted in figure~\ref{figA3to5}.
Fermion coefficients are obtained by $A_n \to (-1)^{n+1} A_n$.

\section{Numerical experiments\label{NumericalExperiments}}

The virial coefficients~(\ref{ExplicitCoefficients}) exhibit
much structure when analysed as function of dimensionality.
They reveal an interesting pattern of zeros. The most prominent feature
is that one zero $d_1(n)$ rapidly approaches
$d=3$ as $n$ increases, and another $d_2(n)$ rapidly approaches
$d=\frac{8}{3}$. Numerically, we find
\begin{equation}
    3-d_1(n) =
   2\cdot10^{-1},
   4\cdot 10^{-2},
   6\cdot 10^{-3},
   7\cdot 10^{-4},\ldots
\end{equation}
for $n=3,4,\ldots$. The convergence turns out to be exponential in $n$,
with some oscillations in the prefactor. Clearly, it is the presence of
the zeros close to $d=3$ which causes the virial expansion to have an unusually large
radius of convergence at $d=3$.

More generally, the set of zeros form an intriguing pattern at low $n$. Investigating this order by order
is an amusing numerical experiment, providing many opportunities to
make (wrong) conjectures. When extending this experiment up to $n=300$, we 
found the most prominent features to be that
i) at $d=2$ all even virial coefficients beyond the
   second vanish, $A_{2n}(2)=0$ for $n=2, 3,\ldots$,
 and ii)
 there are several special dimensions $d_m$ which act
 as ``attractors'' for zeros, with the consequence
 that the virial expansion has an exceptionally large
 convergence radius,
 $\rho_c(d_m) > \rho_{\text{BC}}(d_m)$,
 for such dimensions.

%
\begin{figure}[t]
  \includegraphics[]{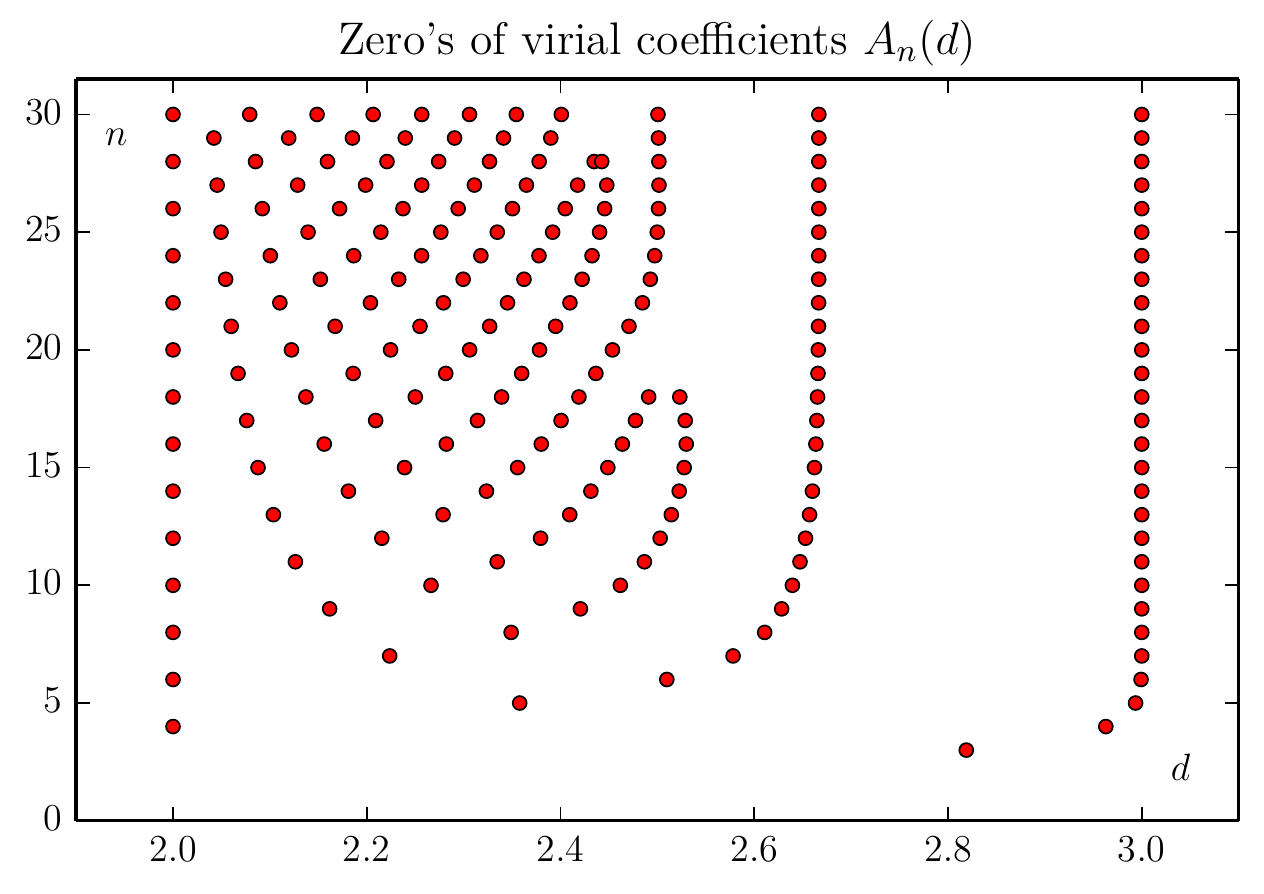}
  \caption{ \label{VirialZerosI} Scatter plot of the zero's of the virial coefficients
  $A_n(d)$ in the region $2\le d$ for $3 \le n \le 30$. There are additional zeros in the region $d < 2$.}
\end{figure}
%

\begin{figure}
  \includegraphics[clip, trim=0ex 0ex 5ex 0ex, width=0.85\linewidth]{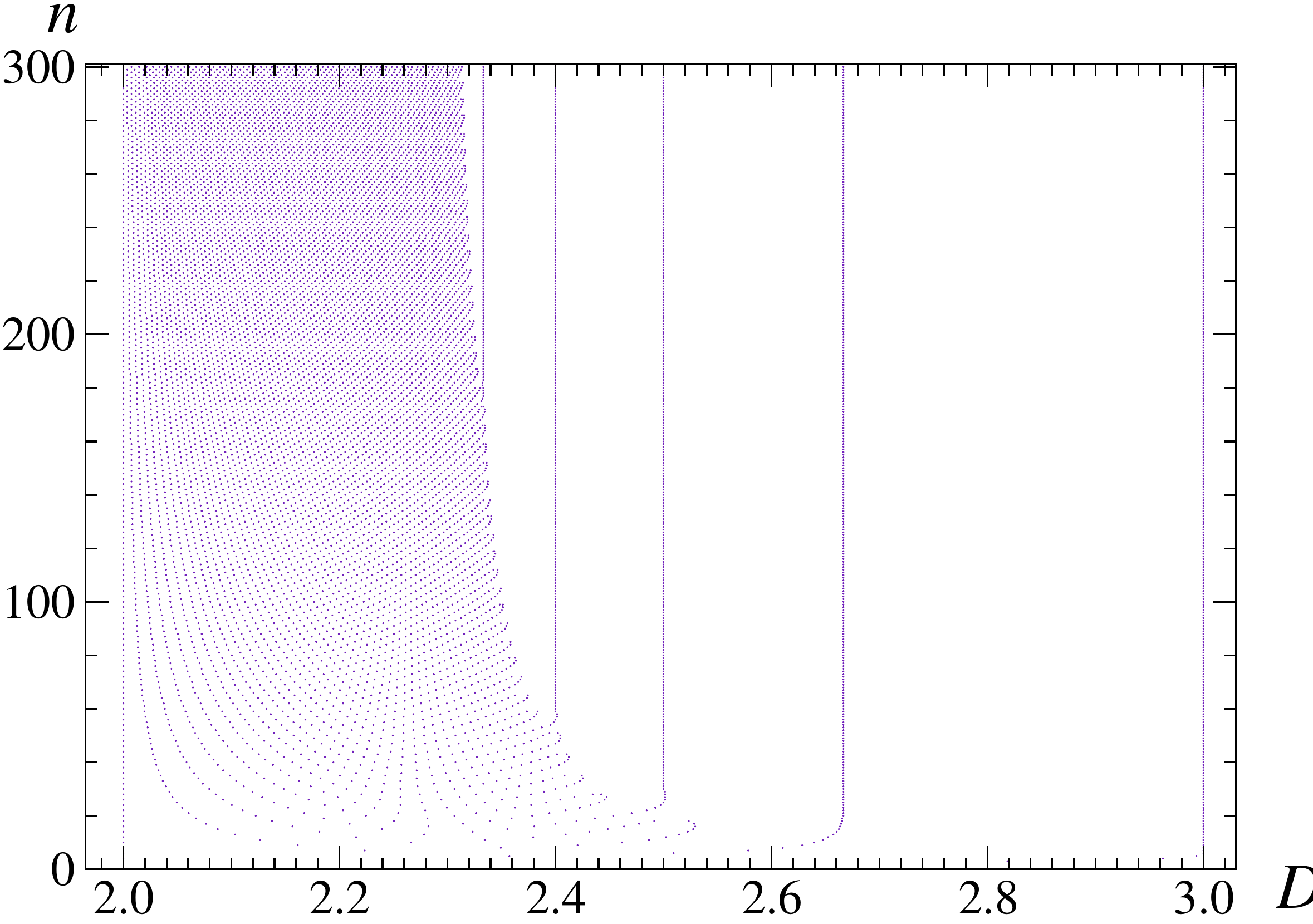}
  \caption{Scatter plot of the zero's of the virial coefficients
  $A_n(d)$ in the region $2\le d$, $3 \le n \le 300$. There are additional
  zeros in the region $d< 2$.}
  \label{VirialZerosII}
\end{figure}

We soon discovered that the special dimensions followed the pattern
\begin{equation}
  d_m = 2 + \frac{2}{m+1}, \quad\text{ $m=1,2,\ldots$,}
  \label{Conjecture}
\end{equation}
leading to the conjecture that this is true for all $m$.
The correct result, which is difficult to discover by
numerical experiments alone, is that equation~(\ref{Conjecture})
holds only for $m=1,2,\ldots,6$. The first 5 members of this set
are clearly visible as the emerging lines in Fig.~\ref{VirialZerosII}.
%
%

\section{Special dimensions\label{SpecialCases}}

We next consider some special cases of $d$. The expressions~(\ref{MayerExpansion}) can be analysed
completely when $d=0$ and $d=2$.

\subsection{Quantum dot}

%
\begin{figure}
\begin{center}
 \includegraphics[]{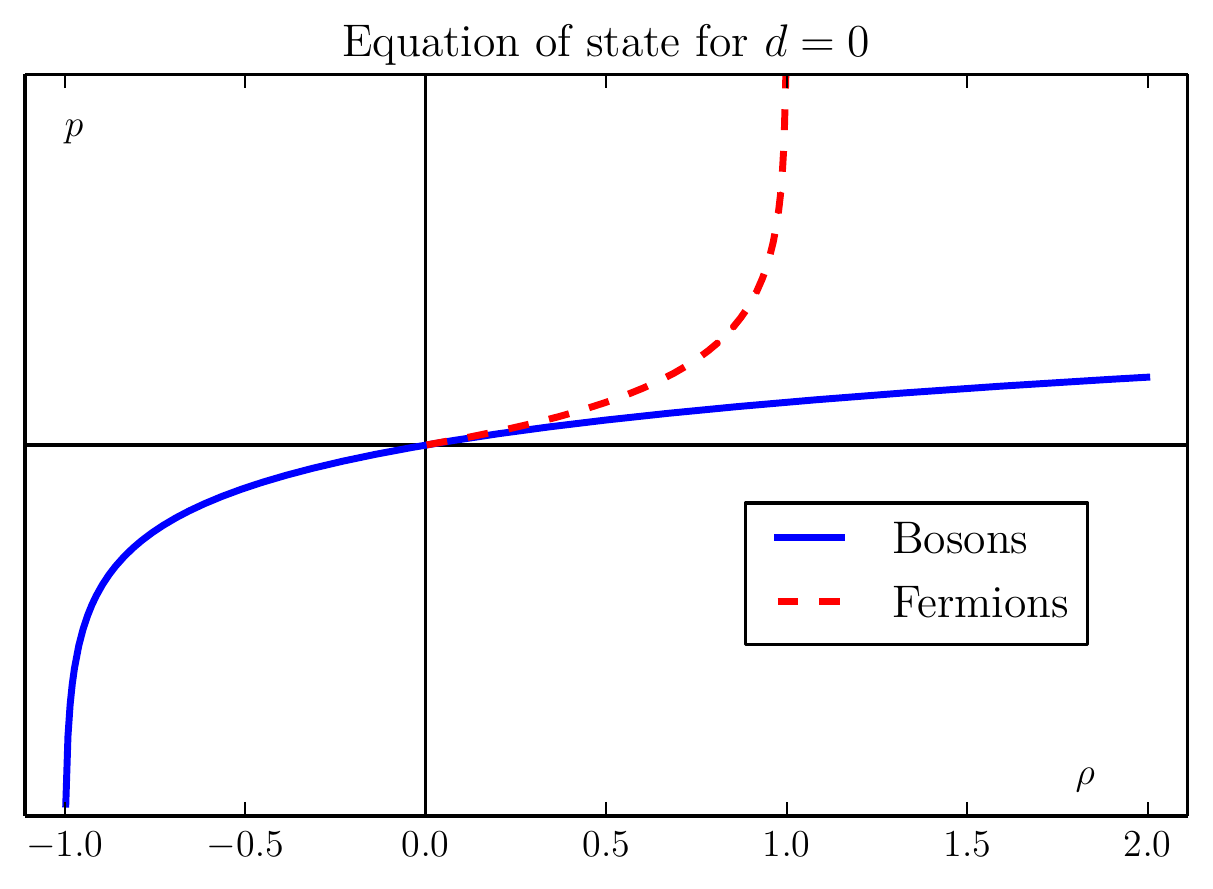}
\end{center}
  \caption{\label{EoS_d0}
    The equation of state for ideal quantum gases in zero dimensions. The boson equation of state in its unphysical 
    third quadrant ($\rho < 0$, $p < 0$) actually describes the fermion equation of state in its physical region, 
    obtained by inversion about the origin as shown. This is a general feature of noninteracting 
    systems in any dimension.
  }
\end{figure}
%

For $d=0$ we find $\rho=z/(1-z)$ and $p=-\ln\left({1-z}\right)$. I.e.
\begin{equation}
   p =\ln\left({1+\rho}\right)= \rho+\sum_{n=2}^\infty \frac{(-1)^{n+1}}{n} \rho^n,
\end{equation}
with virial coefficients $A_n = (-1)^{n+1}/n$ and radius of convergence $\rho_c(0)=1$ due to 
the logarithmic singularity at $\rho=-1$. This singularity has no natural physical interpretation
for bosons. However, it corresponds to the maximum obtainable density for fermions. The $d=0$ 
fermionic equation of state is the same as for a classical hard core lattice gas (without other 
interactions).

\subsection{Two-dimensional layer}

%
\begin{figure}
  \begin{center}
   \includegraphics[]{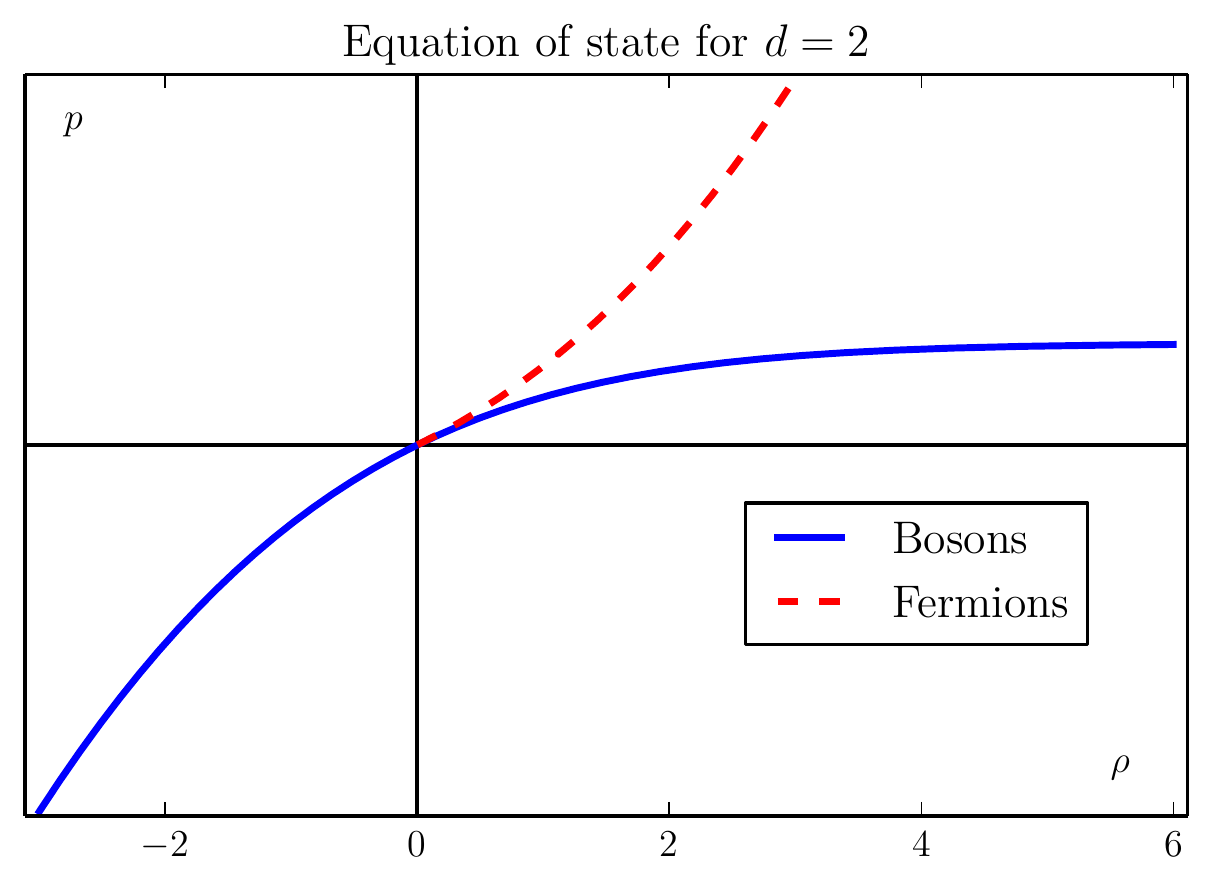}
 \end{center}
  \caption{\label{EoS_d2}
   The equation of state for ideal quantum gases in two dimensions.
  The boson and fermion equations of state only differ in their
  second virial coefficient. Two space dimensions allow for intermediate (anyon) statistics
  which interpolates between the bosons and fermions. The equation of state for such
  hypothetical particles is not exactly known, and probably very far from a smooth
  interpolation at high densities (equivalent to low temperatures).}
\end{figure}
%

For $d=2$, we find $\rho=-\ln(1-z)$ and can infer the diffe\-ren\-ti\-al equation
$(\text{d}p/\text{d}\rho) =\rho/(\text{e}^{\rho}-1)$ with $p(0)=0$. The solution is
\begin{equation}
  p = \int_{0}^{\rho} \frac{\text{d}t\,t}{\text{e}^{t}-1}
  = \text{dilog}(\text{e}^{-\rho}).
\end{equation}
Note that $\text{dilog}(\text{e}^{\rho})=
-\frac{1}{2}\rho^2-\text{dilog}(\text{e}^{-\rho})$,
as can be verified by direct manipulation of the integral.
Thus, the equation of state for fermions and bosons only differ in the second virial coefficient,
$p_{\text{fermion}}=p_{\text{boson}}+\frac{1}{2}\rho^2$. This can also be seen from the
explicit virial coefficients, $A_n = B_{n-1}/n!$ where $B_n$ are the Bernoulli
numbers. Since $B_n=0$ for all odd $n>1$ the virial coefficients $A_n=0$ for all
even $n>2$ (recall that all odd coefficients are equal for bosons and fermions).
The property that only the second virial coefficient depends on statistics seems to generalize to
Haldane exclusion statistics\cite{HaldaneExclusionStatistics} (interpolating between bosons and fermions)
in two dimensions\cite{MurthyShankar}.
The requirements, beyond interpretation of exclusion statistics, are that
i) the density of states is constant as function of energy,
and ii) a position-independent pressure and density can be defined.
The latter condition is not fulfilled by the Hamiltonian model considered in Ref.~\onlinecite{MurthyShankar},
see Ref.~\onlinecite{OuvrydeVeigy}, but this does not
affect their computation of virial coefficients.

Due to pole singularities in the integrand at $t=2\pi\text{i}n$ (with integer 
$n\ne 0$) the function $p(\rho)$ has logarithmic singularities at the same points. 
Hence, the virial expansion has a radius of convergence, $\rho_c(2)=2\pi$. We know 
of no physical reasons for the occurence of these singularities (contrary to the 
$d=0$ case, where the singularity has a physical origin in the fermion system).

For later analysis, recall that a logarithmic singularity has infinitely
many Riemann sheets. In this case, when $\rho$ encircles the singularity at 
$2\pi\text{i}m$ once in the clockwise direction the function changes by 
$p \to p -4\pi^2 m$. The result of repeated encirclings is that 
\begin{equation}
  p \to p -4\pi^2 \sum_m m\,k_m
\end{equation}
when the singularity at $\rho = 2\pi\text{i}m$ is encircled $k_m$ times in total, counted in
the clockwise direction. The order in which the encirclings occur does not matter. Hence, in 
the two-dimensional ($d=2$) case, each Riemann sheet is uniquely labeled by a single integer 
$N = \sum_m m\,k_m$. The total surface is multiply connected, so there are infinitely many 
topologically inequivalent ways of moving from one sheet to another. Amusingly, the {\em same\/}
Mayer expansion (\ref{MayerExpansion}) (for $d=2$) defines an infinity of possible equations 
of state, depending on which Riemann sheet is selected. All these equations of state are physical 
in the sense that the pressure is real when the density is real.

As will be seen in Section \ref{RiemannSurface} this behavior extends to all dimensions 
$d$, with the generalization that each Riemann sheet is labeled by an infinite-dimensional vector 
$\bm{k}$ of integers, which generically must be restricted by the condition $k_m = -k_{-m}$ for 
the pressure to be real when the density is real.

\subsection{Three dimensions}

%
\begin{figure}
  \includegraphics[]{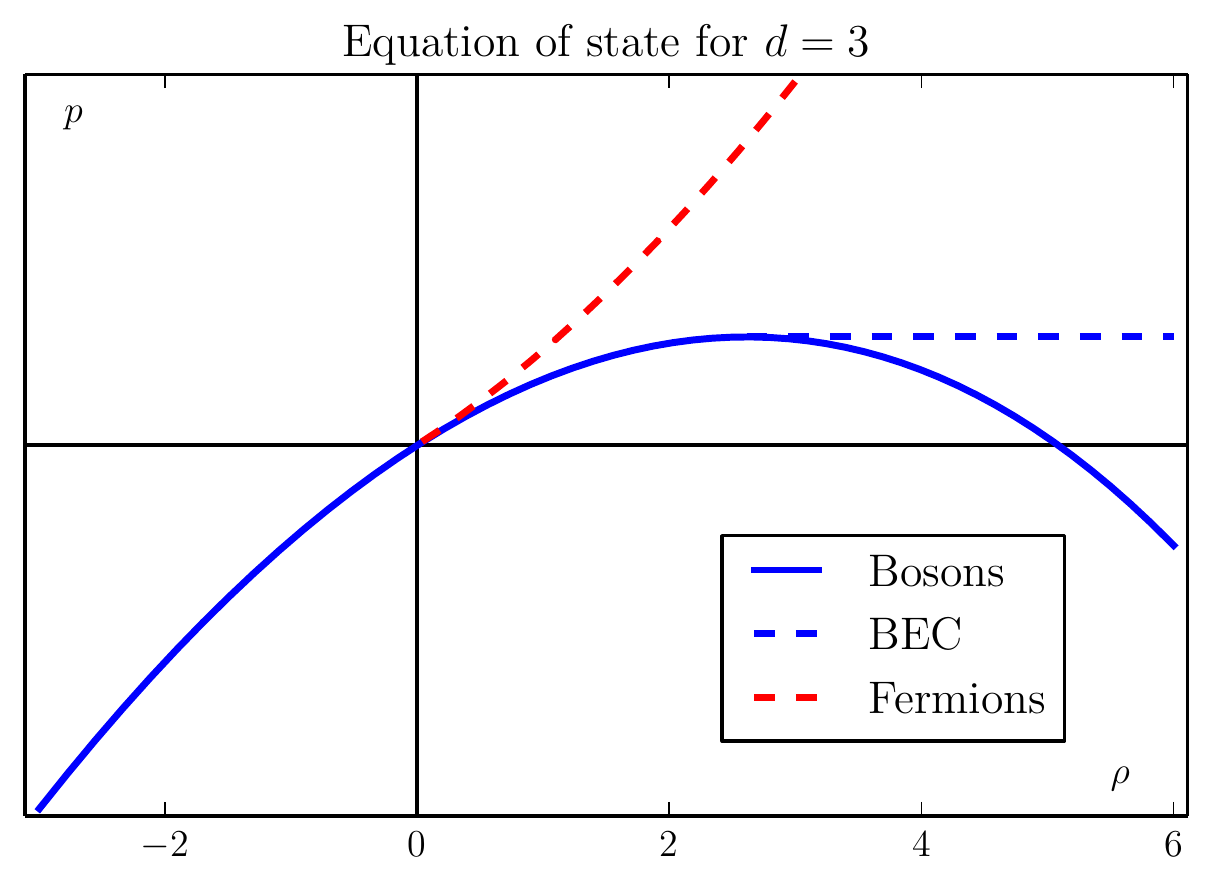}
  \caption{\label{EoS_d3}
  The equation of state for ideal quantum gases in three dimensions. The pressure
  as a function of density can be analytically continued beyond the critical density for 
  Bose-Einstein condensation, $\rho_{\text{BE}}=\zeta(\frac{3}{2})=2.612\ldots$. The
  same is true for other thermodynamic quantities, like the chemical potential 
  $\mu=\mu(\rho)$. However, the behavior for $\rho > \rho_{\text{BE}}$ is unphysical 
  (with $(\text{d}p/\text{d}\rho) < 0$ and $(\text{d}\mu/\text{d}\rho) < 0$), and does 
  not correspond to the correct infinite volume limit of the system.
  }
\end{figure}
%

In three dimensions, it seems impossible to invert
the Mayer expansion~(\ref{MayerExpansion}) explicitly.
We have evaluated $A_n(3)$ numerically to high accuracy
and quite large $n$. They behave like
\begin{equation}
   A_n(3) \sim \mathcal{A}(n)\, \exp\left(-a n \right)\cos(bn+c)
\end{equation}
when $n$ becomes large. Here $\mathcal{A}(n)$ changes quite slowly and smoothly
(i.e.\ algebraically) with $n$, and $a\approx 2.9$, $b\approx 0.7$.
As shown in Fig.~\ref{EoS_d3} an equation of state can be computed
from this virial expansion far beyond the Bose-Einstein
condensation point $\rho_{\text{BE}}=\zeta({\frac{3}{2}})\approx 2.612\ldots$. 
However, the solid curve is unphysical  for $\rho > \rho_{\text{BE}}$, since pressure decreases 
with increasing density.

In general, the virial coefficients can be expressed by a contour integral
\begin{equation}
  A_n = \int_{\mathcal{C}} \frac{d\rho}{2\pi i}\,\frac{p(\rho)}{\rho^{n+1}}.
  \label{Coefficient_representation}
\end{equation}
For evaluation at large $n$ one should deform the closed curve $\mathcal{C}$ as far away 
from the origin as possible, since the behavior of $A_n$ then is governed by the contributions 
from the nearest singularities of $p(\rho)$, beyond which $\mathcal{C}$ cannot be deformed. With a 
pair of complex conjugate singularities, at $\rho_+=\rho_c\,\text{e}^{\text{i}\omega}$ and 
$\rho_-=\rho_c\,\text{e}^{-\text{i}\omega}$, one obtains an asymptotic behavior
\begin{equation}
    A_n \sim \rho_c^{-n}\,\cos\left(n\omega + \phi \right),
    \label{IntegralExpression}
\end{equation}
up to algebraic corrections in $n$. The phase $\phi$ is not important for locating the 
singularities. By comparing this result with the numerical observations, one concludes that 
the convergence of the $d=3$ virial expansion is governed by a complex conjugate pair of
singularities at $\rho_\pm \approx \text{exp}({2.9\pm 0.7\text{i}})
\approx 18.5\, \text{exp}({\pm 0.7\text{i}}) \approx 14 \pm 12\text{i}$.
This is in agreement with earlier findings that the radius of convergence is much larger 
than the critical density for Bose-Einstein condensation. We want to study these singularities 
in more detail.

\section{\label{RiemannSurface} Analytic continuation
and Riemann surface}

There are two possible sources of singularites in an implicitly represented function $p(\rho)=p(z(\rho))$, namely
i) singularities which are explicit in the parametric representation of $p=p(z)$ or $\rho=\rho(z)$, or both (in which case they 
may sometimes compensate each other), and ii) singularities occurring where $\rho(z)$ is analytic, but cannot be inverted to an 
analytic function $z(\rho)$ because $\text{d}\rho(z)/\text{d}z=0$. The singularities of relevance to our case are 
of the second kind. To find their accurate positions in the $\rho$-plane, and the corresponding $z$-plane values, one must extend 
the parametric representation (\ref{ImplicitEquationOfState}) to the full Riemann surface of the functions involved.

The expressions involved are known as polylogarithmic functions,
\begin{equation}
    \text{Li}_s(z) =
    \frac{1}{\Gamma(s)}\int^\infty_{0} \frac{\text{d}\varepsilon}{\varepsilon}\,
    \frac{z\,\varepsilon^s}{\text{e}^{\varepsilon}-z}
    = \sum_{n=1}^{\infty} \frac{z^n}{n^s}.
\end{equation}
The sum converges for $\vert z \vert < 1$ and has an analytic extension equal to the integral expression in 
the whole $z$-plane cut along the real $z$-axis from 1 to $\infty$. This is the primary Riemann sheet of the 
polylogarithm. Moving $z$ along a closed path which winds once clockwise around $z=1$ (not encircling $z=0$) 
changes the integral so that
\begin{equation}
    \text{Li}_s(z) \to \text{Li}_s(z) +
    \frac{2\pi\text{i}}{\Gamma(s)} \left(\log z\right)^{s-1},
  \label{OnceAround}
\end{equation}
where the right hand side again is defined on the primary Riemann sheet. The new term arises because moving 
$z$ across the positive real $\varepsilon$-axis drags the integration path with it. This can be undone by 
explicitly evaluating the pole contribution which makes up the difference. The new term must be handled 
carefully, since $\log z$ is multivalued around $z=0$ and $\left(\log z\right)^{s-1}$ is multivalued 
around $z=1$. We introduce $\mu\equiv \log z$ as a new variable, in terms of which the singularity at $z=1$ 
becomes the image of infinitely many singularities in the $\mu$-plane, at $2\pi\text{i}m$, $m=0, \pm1, \pm2, \ldots$. 
We define the primary Riemann sheet by introducing cuts $2\pi\text{i}m+x$, $0\le x\le \infty$. If we start
on the primary sheet and make $k$ crossings (counted in the clockwise direction) of the cut from $2\pi\text{i}m$ 
to $\infty$, and no crossings of any other cuts, the polylogarithm will change as
\begin{align}
    \text{Li}_s(\text{e}^\mu) &\to  \text{Li}_s(\text{e}^\mu)\label{ContinuedFunction}\\
    &+ \Gamma(1-s)\left(\text{e}^{2\pi\text{i}k(1-s)}-1\right)\left(2\pi\text{i}m-\mu\right)^{s-1}.\nonumber
\end{align}
This is found most safely from the expression\cite{Bateman_Project},
\begin{equation}
  \sum_{n=1}^\infty\,\frac{\text{e}^{n\mu}}{n^s} =
  \Gamma(1-s)\,(-\mu)^{s-1} +
  \sum_{n=0}^\infty \zeta(s-n)\,\frac{\mu^n}{n!},
  \label{BatemanExpression}
\end{equation}
which is convergent in a finite region around $\mu=0$. It can also be seen directly, in that each new crossing adds a 
new function, but also changes the phase of the previously emerged ones. Hence, the total contribution from $k$ 
crossings is
\begin{align*}
   &\frac{2\pi\text{i}}{\Gamma(s)} \sum_{n=0}^{k-1}
   \text{e}^{-2\pi\text{i}n(s-1)} \left(\ln z\right)^{s-1}\\
   &= \frac{\pi}{\Gamma(s)\,\sin \pi(1-s)}
   \left(\text{e}^{2\pi\text{i}k(1-s)}-1\right) \left(\text{e}^{-\pi\text{i}} \ln z\right)^{s-1}.
\end{align*}
By paying close attention to phase relations on the primary sheet one finds
$\text{e}^{-\pi\text{i}} \ln z = -\mu$. By further using the formula 
$\Gamma(t)\Gamma(1-t)=\pi/\sin\pi t$ one finds agreement with equation~(\ref{ContinuedFunction}). 
This equation is also valid if $k$ is negative, that is if one makes crossings in the anticlockwise
direction.

The contributions  from crossing different branch cuts are additive and do not interfere with each 
other. Thus, it does not matter in which order the various crossings have occurred, only their total 
number. This is a great simplification. Thus, we may label the Riemann sheets by an integer-valued 
vector $\bm{k}$, where $k_m$ is the net number of times the branch cut from  $2\pi\text{i}m$ to $\infty$ 
has been crossed in the clockwise direction (starting from the primary sheet), and define a polylogarithm 
extended to the complete Riemann surface as
\begin{equation}
  \text{Li}(\mu,s;\bm{k}) = \text{Li}_s(\text{e}^\mu) +
  \sum_{m=-\infty}^{\infty} C(k_m,s)\,\left(2\pi\text{i}m-\mu\right)^{s-1},
\end{equation}
where
\begin{align}
    C(k,s) &= \frac{2\pi\text{i}}{\Gamma(s)}\,
    \frac{\sin\pi k(1-s)}{\sin\pi(1-s)}\,\text{e}^{\pi\text{i}k(1-s)}.
\end{align}
When starting from the primary sheet, in practice only a few (if any) of the coefficients $k_m$ will be 
nonzero. Equation~(\ref{ContinuedFunction}) provides the analytically continued expressions for pressure 
and density,
\begin{align}
    p    &= \text{Li}(\mu,1+d/2;\bm{k}),\nonumber\\[-0.7ex]
    \phantom{\cdot}\label{GeneralEoS}\\[-0.7ex]
    \rho &= \text{Li}(\mu,d/2;\bm{k}).\nonumber
\end{align}
Note that $C(-k,s)=C(k,s)^*$. Therefore, $p$ and $\rho$ will be real when $k_m = -k_{-m}$ for all $m$, 
thereby making the relation between $p$ and $\rho$ a candidate for a physical bosonic equation of state (with
the physical region defined by $\mu$ being real). Additional possibilities arise for rational values
of $s$, when the logarithmic singularities at $\mu=2\pi\text{i}m$ become algebraic and the coefficients 
$C$ become periodic in $k$. Note that
\begin{align*}
   C(k,m)   &= (-1)^{m-1} \frac{2\pi\text{i}k}{(m-1)!} &\quad\text{for $m=1,2,\ldots$,}\\
   C(k,-m)  &= 0 &\quad\text{ for $m=0,1,\ldots$.}
\end{align*}

\section{Locating the singularites\label{TravelAccount}}

It would be almost impossible to find the singularities of interest if one did not know where to look.
Fortunately, we are in a situation quite similar to the one of explorers searching for a magnetic
pole, in that we have a compass showing which direction to follow.

%
\begin{figure}
  \includegraphics[width=0.8\linewidth]{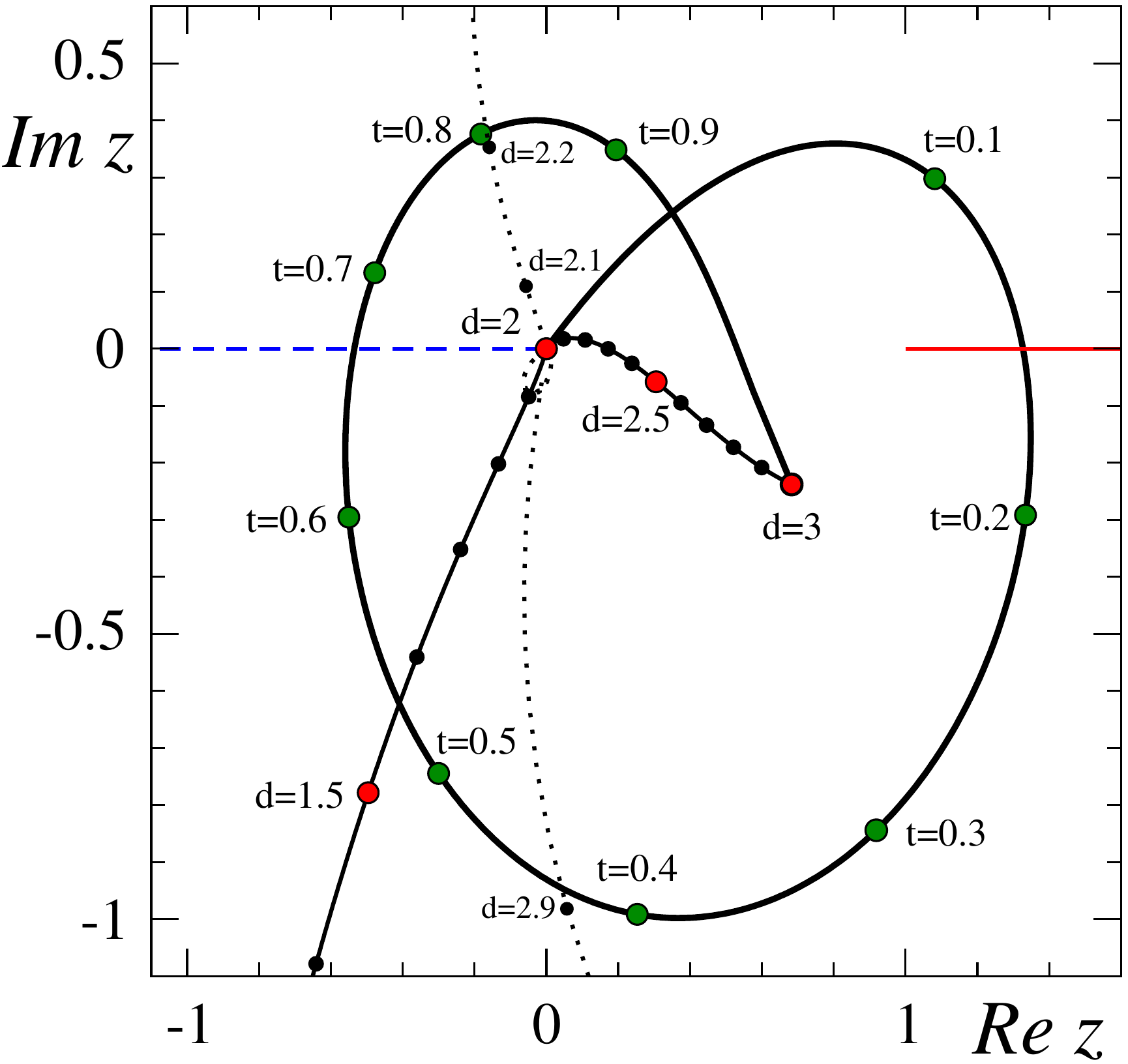}
  \caption[]{\label{FindingRhoPlus}\linewidth=1.1\linewidth
  This figure maps our explorations of singularities. The fulldrawn line (colored red) is the first branch cut,
   running from $z=1$ to $z=\infty$ along the positive real axis. The dashed line (colored blue) is the
   second branch cut, running from $z=-\infty$ to $z=0$ along the negative real axis. We first found the (fulldrawn 
   black) line in the $z$-plane which corresponds to the curve $\rho(t)=t\,\rho_+(3)$, $0\le t\le1$,
   starting at $z=0$. The filled circles (colored green)
   mark the progression as $t$ increases. As can be seen,
   one must cross two branch cuts before reaching the singularity $\rho_+(3)$.\\[0.5ex]
   Having found $\rho_+(3)$ and equation (\ref{Second_Singularity})
   determining $\rho_+(d)$, it is easy to explore how the singularity moves when one changes $d$.
   This is shown by the next fulldrawn (black) line.
   The black dots and filled circles (colored red) mark
   the progression as $d$ decreases. When $d\to0$ the singularity moves
   to $z=\infty$ (multiplied
   by a phase), corresponding to a singular point at $\rho=-1$. Actually,
   (\ref{Second_Singularity}) becomes ambiguous when $d\to 2, 0, -2,\ldots$.
   To handle this, we may write $d=2+\varepsilon\,\text{e}^{\text{i}\varphi}$
   with $\varepsilon$ a small positive number ($0.001$), and
   varied $\varphi$ from $0$ to $\pi$. In the $z$-plane the singularity moves
   slightly above $z=0$ and back across the second branch cut. Alternatively, one may
   vary $\varphi$ from $0$ to an infinite number of other possible endpoints $(2k+1)\pi$.
   Each choice leads to a different solution. There are infinitely many singularites approching 
   $\rho=2\pi\text{i}$ as $d$ approaches $2$. We have mapped out several endpoint choices
   before decreasing $d$ to $0$ or increasing $d$ back to $3$. The two black dotted lines show 
   how the singularity moves when the encircling is chosen  to $\pm2\pi$, and $d$ is increased 
   back to $3$. See the caption to Fig.~\ref{ChangingDimension} for a  further description.}
\end{figure}
%

From the numerically calculated virial coefficients for $d=3$, we conclude that the behavior is governed by the singularity at $\rho_{+}\approx 14 + 12\text{i}$
and its complex conjugate. It is then easy to map out a path from the origin to $\rho_+$ by solving the equation $\rho(\mu(t))=t\rho_{+}$ numerically and step 
by step for $t=\Delta t, 2\Delta t,\ldots$. This leads to the Riemann sheet defined by $k_0=1$, all other $k_m=0$.  As $t$ approaches 1, the inversion
becomes difficult in accordance with our expectation that  $\rho(z)$ cannot be inverted at the singularity, namely
\begin{equation}
    \frac{\text{d}\rho}{\text{d}z} = \frac{1}{z}\, \text{Li}(\mu,-1+d/2;\bm{k})=0.
    \label{Second_Singularity}
\end{equation}
One may then change the search algorithm to a direct solution of this equation, which is explicitly
\begin{equation}
  -\sqrt{\frac{4\pi}{-\mu}} +
  \sum_{n=1}^{\infty} \frac{\text{e}^{n\mu}}{n^{1/2}} =0.
  \label{SecondSingularity}
\end{equation}
This equation can be solved numerically to essentially aribitrary accuracy. We find
\begin{align}
   \mu_+  &=  -0.322\,995\,155\,543\,097
   \;-\; 6.618\,613\,424\,959\,805\;\text{i},\nonumber\\
   z_+    &=  \phantom{-}0.683\,629\,720\,405\,578
   \;-\; 0.238\,314\,132\,061\,880\;\text{i},\nonumber\\
   \rho_+ &=  \,\,14.074\,421\,676\,564\,36 \;+\; 
   12.107\,496\,215\,707\,89 \;\text{i}\nonumber\\
           &= \,\,18.565\,581\,330\,600\,061
           \;\text{e}^{0.710\,413\,678\,806\,621\;\text{i}}.
           \label{Accurate_rho_c}
\end{align}
The results of Ref.~\onlinecite{Jenssen_Hemmer} and \onlinecite{Ziff} are in agreement with (\ref{Accurate_rho_c}) to within the 
estimated accuracy.

\begin{figure}
  \includegraphics[width=0.77\linewidth]{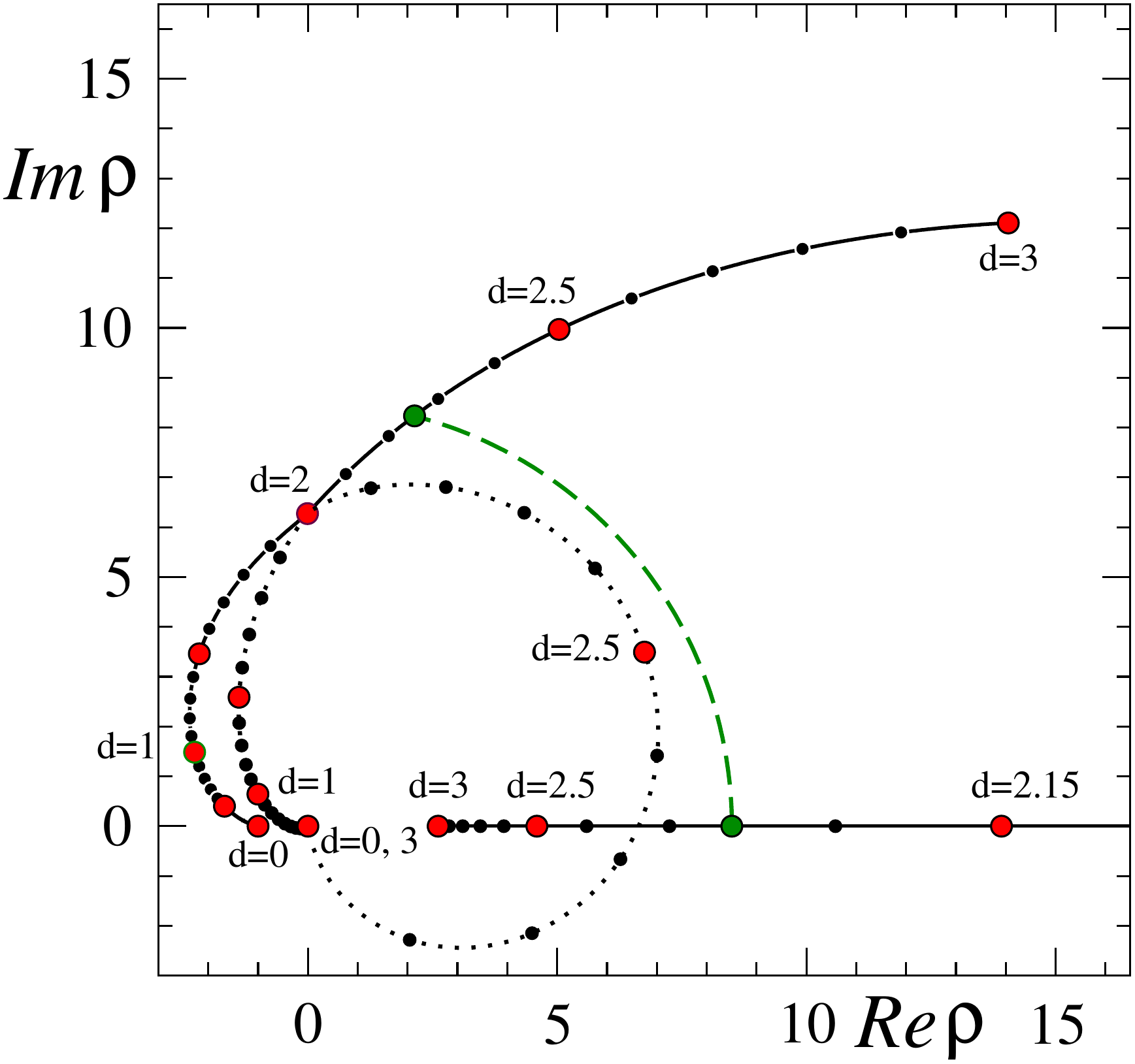}
  \caption{This figure shows how the singularities $\rho_+(d)$ and
  $\rho_{\text{BE}}(d)$ move as one changes the dimension $d$. Starting near $d=3$ the
  singularity $\rho_{\text{BE}}(d)$ is closer to the origin than $\rho_+(d)$, thus
  determining the convergence radius of the virial expansion (except when $d$ is equal to
  one of the special points $d_m$).
  However, as $d$ decreases $\rho_+(d)$ moves inwards while $\rho_{\text{BE}}(d)$ moves
  outwards. Thus, there is a crossover dimension $d_c$,
  indicated by the filled circles (colored green),
  where they are equally far from the origin (as indicated by the
  green dashed circle). We find $d_c\approx 2.252\,563\,996$.
  For $d<d_c$ the convergence radius of the virial expansion is always 
  determined by $\rho_+(d)$ (and its complex conjugate $\rho_-(d)$). Since
  ${18}/{8}< d_c < {16}/{7}$ the sequence of special points $d_s$ ends
  at $d_s=16/{7}$.
  As we decrease $d$ further towards $2$ one discovers that $d=2$ is a
  logarithmic singular point for $\rho_+(d)$. Hence, one may encircle
  $d=2$ in various ways before changing $d$ further. 
  The figure displays the two behaviors when $d$ is decreased
  to $0$ (after encircling $d=2$ by angles $\pm\pi$). In one case
  the singularity moves to $\rho=-1$ when $d\to0$, which is a
  logarithmic singularity of the $d=0$ equation of state.
  In the other case $\rho_+$ moves to $\rho=0$ when $d\to0$. 
  There is no singularity at $\rho=0$ in the $d=0$ equation
  of state, but it is possible for two square root singularities to
  annihilate when they coalesce. Finally, the figure shows the behavior
  when $d=2$ is encircled by $2\pi$, and increased back to $d=3$.
  In this case $\rho_+$ move to $\rho=0$, where it appears to
  annihilate with $\rho_-$.  Thus, the analytically continued equation of
  state has singularities close to $\rho=0$ when $d$ is close to
  $3$ and $0$, but not for $d$ {\em exactly\/} $3$ or $0$.  
}
  \label{ChangingDimension}
\end{figure}

A $z$-plane map of the journey described above, to the singularity at $\rho_+$, is shown in Fig.~\ref{FindingRhoPlus} and described 
in the figure caption. The other journey follows the complex conjugate path. As shown, one has to cross two branch cuts in the complex 
$z$-plane. When the singularity first has been located it is very easy to follow its movement as one changes the dimension $d$. We 
discuss this in more detail in the captions to Fig.~\ref{FindingRhoPlus} and Fig.~\ref{ChangingDimension}, and further in Section
~\ref{ExploringDimension}.

%
%

\section{Asymptotic virial coefficients\label{AsymptoticCoefficients}}

With accurate information on the singular behavior we may use~(\ref{Coefficient_representation}) to compute the asymptotic behavior 
of $A_n$ as $n\to\infty$. We may expand $\rho(\mu)$ and $p(\mu)$ around $\mu_+$, that is, with $\Delta\mu = \mu-\mu_+$, $\Delta\rho 
= \rho_+-\rho$, and $\Delta p = p-p_+$, where
\begin{align}
    \Delta\rho =  -\sum_{n=2}  \frac{r_n}{n!}\Delta\mu^n,\quad
    \Delta p     = \sum_{n=1} \frac{p_n}{n!} \Delta\mu^n.
\end{align}
Here, $p_n = r_{n-1} = \text{Li}(\mu_+,1-n+d/2; \bm{k})$.
In particular, $p_1 = \rho_+$ and $p_2 = r_1 = 0$ (the singularity condition).
We may now express $\Delta\mu$ order by order in $\Delta\rho$, thus
\begin{align}
  \Delta\mu &= ({-2}/{r_2})^{1/2}\,\sqrt{\Delta\rho} + \cdots,\nonumber\\
  \Delta p    &=  ({-2}/{r_2})^{1/2}\,\rho_+\,\sqrt{\Delta\rho} +
  \cdots \label{p_singularity}
\end{align}
We insert this expansion for the pressure into the integral expression
(\ref{Coefficient_representation}), and deform the integration
contour around the square root branch cut starting at $\rho=\rho_+$.
Writing $\rho = \rho_+(1+t)$ gives a contribution
\begin{align*}
    A^{(+)}_n &=
    \frac{1}{\pi}\,(2/r_2)^{1/2}\,\rho_+^{{3/2}-n}\,
    \int_0^\infty  \frac{\text{d}t \,t^{1/2}}{(1 + t)^{n+1}} + \cdots\\
    &=
    \frac{1}{\sqrt{2\pi\, r_2}}\,\rho_+^{3/2-n}\,n^{-3/2} + \cdots
\end{align*}
to~(\ref{Coefficient_representation}).  There is a similar contribution $A_n^{(-)} = A_n^{(+) *}$ from
the singularity at $\rho_-$. There are also higher order contributions from weaker singularities (proportional 
to $\Delta\rho^{m+1/2}$) in $\Delta p$.  This leads to the following expansions
{\footnotesize
\begin{align}
  A_n  &\approx \vert \rho_+ \vert^{3/2-n}\,\sum_{m=0}
  c_m\,\cos\left(\omega\,n +
    \phi_m\right)\,\int_{0}^{\infty}\frac{\text{d}t\,t^{1/2+m}}{\left(1+t\right)^{n+1}}\nonumber\\
  &= n^{-3/2}\,\vert \rho_+ \vert^{3/2-n}\,\sum_{m=0}
     n^{-m}\,a_{m}\,\cos\left(\omega\,n +
       \alpha_m\right),\label{AsymptoticExpansion}
\end{align}
}with amplitudes $a_m$, $c_m$ and phases
$\alpha_m$, $\phi_m$ 
which are straightforward to compute. The first numerical values of $a_m$ and $\alpha_m$ for dimension $d=3$ are listed in
table~\ref{VirialCoefficientsAsymptoticExpansion}.

\begin{table}
\begin{center}
{\footnotesize
\begin{tabular}{|c|r|r|}
\hline
$m$&\multicolumn{1}{c|}{$a_m$}&\multicolumn{1}{c|}{$\alpha_m$}\\
\hline
${0}$&${0.489\,092\,994\,674\,599}$&${-0.115\,262\,558\,466\,782}$\\
${1}$&${0.807\,395\,011\,652\,565}$&${-0.639\,351\,821\,397\,655}$\\
${2}$&${3.094\,390\,113\,185\,892}$&${-1.159\,038\,217\,120\,639}$\\
${3}$&${17.394\,603\,934\,793\,184}$&${-1.466\,809\,677\,404\,770}$\\
\hline
\end{tabular}
}
\end{center}
\vspace{1ex}
\caption{\label{VirialCoefficientsAsymptoticExpansion} 
Numerical parameters for the asymptotic expansion of virial coefficients $A_n$ in $3$ dimensions.
Equation~(\ref{AsymptoticExpansion}), with
$\vert \rho_+ \vert = \exp\left(2.921\,309\,400\,394\,924\right)$
and $\omega = 0.710\,413\,678\,806\,621$, provide an accurate
representation for large $n$. The relative error in $A_n$ when summing
the series in (\ref{AsymptoticExpansion}) to $m=k$ can be expected to
be of order $n^{-(k+1)} (a_{k+1}/a_0)$.
}
\end{table}

%
\begin{figure}
\begin{center}
\includegraphics[]{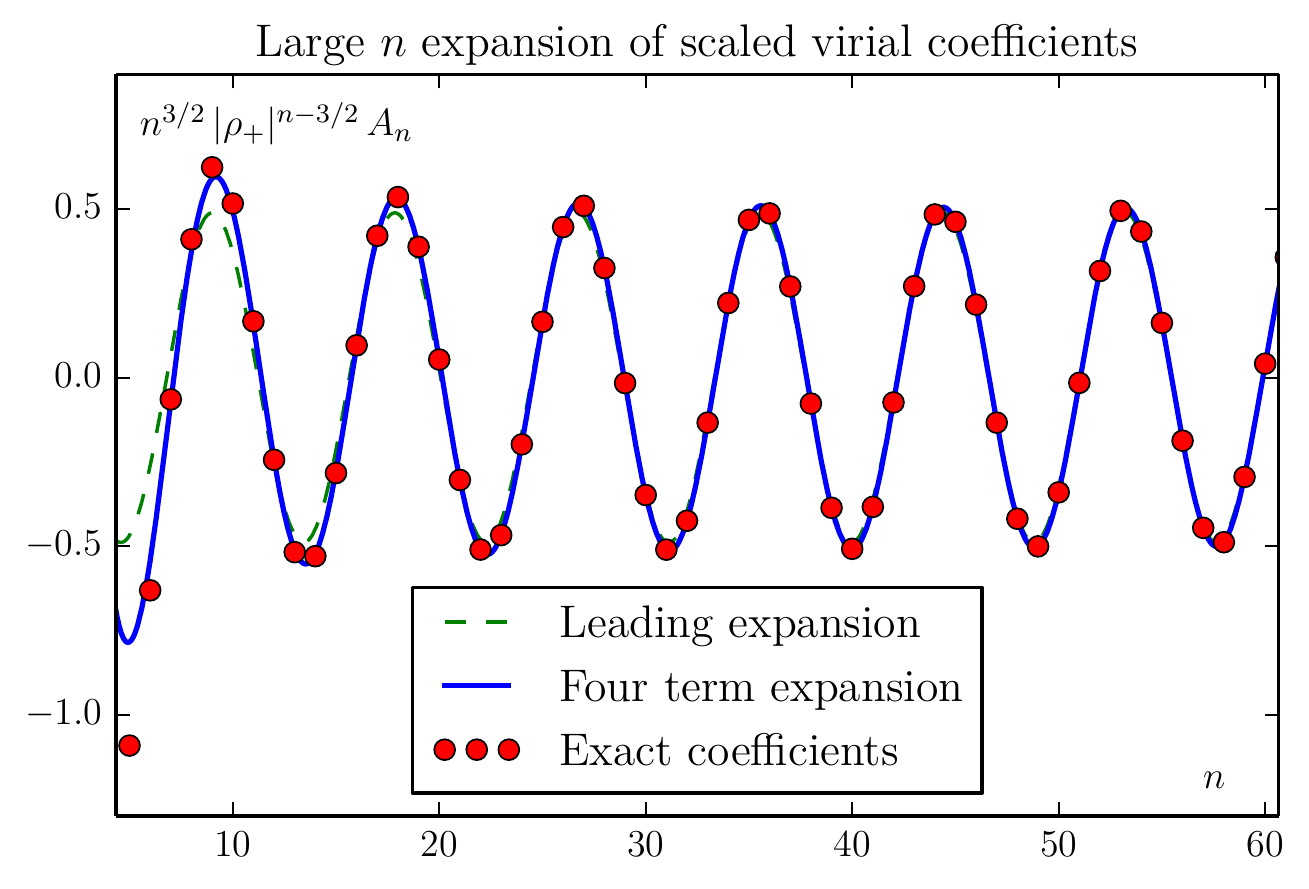}
\end{center}
\caption{\label{asymptoticAn}The exactly
computed values of scaled virial coefficients compared with
the expansion~(\ref{AsymptoticExpansion}). The expansion
does not work well for low $n$, due to contributions from additional
singular points (further away from the origin) of the equation of state.
}
\end{figure}
%

Because of the oscillatory behavior, it is difficult to find an accurate asymptotic fit to $A_n$ directly
from the numerical series, in particular if the prefactor $n^{-3/2}$ is unknown. In retrospect one
realizes that this $n$-dependence is the generic behavior of the algebraic prefactor, due to the
square root type of the singularity~(\ref{p_singularity}).

\section{Behavior at the Bose--Einstein condensation point\label{NearBEC}}

To understand why the dimensions $d_m$ are special for the virial expansion, with a radius of convergence much larger
than $\rho_{\text{BE}}$, we must investigate the equation of state for $\rho$ near $\rho_{\text{BE}}$. The formula~(\ref{BatemanExpression}) is useful for this.
We restrict analysis to the interval $2< d < 4$.
It follows that
\begin{align*}
  \left( \rho_{\text{BE}}-\rho\right)
  &= -\Gamma(1-\frac{1}{2}d)\,\left(-\mu\right)^{-1+d/2}+\cdots,\\
  \left( p_{\text{BE}}-p\right)
  &= \zeta(\frac{d}{2})\,\left(-\mu\right)+\cdots,
\end{align*}
as $\mu\to 0^-$, with $\rho_{\text{BE}}=\zeta(\frac{d}{2})$
and $p_{\text{BE}}=\zeta(1+\frac{d}{2})$.
Elimination of $\mu$ gives the equation of state as $\rho\to\rho_{\text{BE}}$ from below:
\begin{equation}
  \left( p_{\text{BE}}-p\right) =
  \rho_{\text{BE}} 
  \left[\frac{\left(\rho_{\text{BE}}-\rho\right)}{-\Gamma(-1/\delta)}\right]^{\delta}+\cdots,
  \label{Critical_EoS}
\end{equation}
where $\delta=2/(d-2)$. This term is non-singular when $\delta$ is equal to an integer $1+m > 1$. This corresponds to dimensions 
$d=2+2/(m+1)$ with $m=1,2,\ldots$, giving $p_{\text{BE}}-p \sim \left(\rho_{\text{BE}}-\rho\right)^{m+1}$.  For odd $m$ this equation 
of state is obviously unphysical when $\rho > \rho_{\text{B}}$, but this is not equally obvious when $m$ is even. There is, however, 
a signature of singular behavior in the density fluctuations, which diverge like $\left(\rho_{\text{BE}}-\rho\right)^{-m}$.

One must also check that there are no singularities to higher orders. For this it follows from equation~(\ref{BatemanExpression})
that $p$ and $\rho$ are analytic functions of $\xi \equiv (-\mu)^{1/\delta}$ in a region around $\mu=0$ (the power series converges), 
with $d\rho/d\xi\ne0$ at $\xi=0$. Therefore, the function $\rho(\xi)$ can be inverted. This implies that $\xi$, $\mu=-\xi^\delta$, 
and $p$ are analytic functions of $\rho$ in some region around $\rho_{\text{BE}}$ for all special dimensions given by 
equation~(\ref{Conjecture}). However, this does not necessarily imply that there will be a sudden increase of the
convergence radius of the virial expansion at these dimensions, since the convergence may be governed by singularities
closer to the origin than $\rho_{\text{BE}}$.

Returning to general dimensions, $2< d <4$, we insert equation~(\ref{Critical_EoS}) into (\ref{Coefficient_representation}) to obtain the 
asymptotic contribution to the virial coefficients from the Bose-Einstein singularity. By deforming the integration contour around the 
branch cut starting at $\rho=\rho_{\text{BE}}$ and performing the integration, we find 
\begin{equation}
     A^{\text{(BE)}}_n(d) =
     \mathcal{B}(d)\,B(1+\delta,n-\delta)\,\rho_{\text{BE}}^{1+\delta-n} + \cdots
     \label{BoseEinsteinContribution}
\end{equation}
as ${n\to\infty}$. Here $B(1+\delta,n-\delta)$ is the beta function, it behaves like $n^{-(1+\delta)}$ as $n\to\infty$. The
coefficient in front is
\begin{equation}
    \mathcal{B}(d) = \frac{1}{\pi}\,\left[-\Gamma(1-{d}/{2})\right]^{-\delta}\,\sin(\pi\delta).
\end{equation}
The Gamma function has a pole singularity as $d\to4$ (i.e.~$\delta\to1$), cancelling the zero in $\sin\pi\delta$. Thus, the equation of 
state is nonanalytic at $\rho_{\text{BE}}$ for $d=4$ (which corresponds to $m=0$ in equation~(\ref{Conjecture})). This can also be seen 
by direct analysis of the equation of state for $d=4$.
 
\begin{figure}
  \includegraphics[width = 0.85\linewidth]{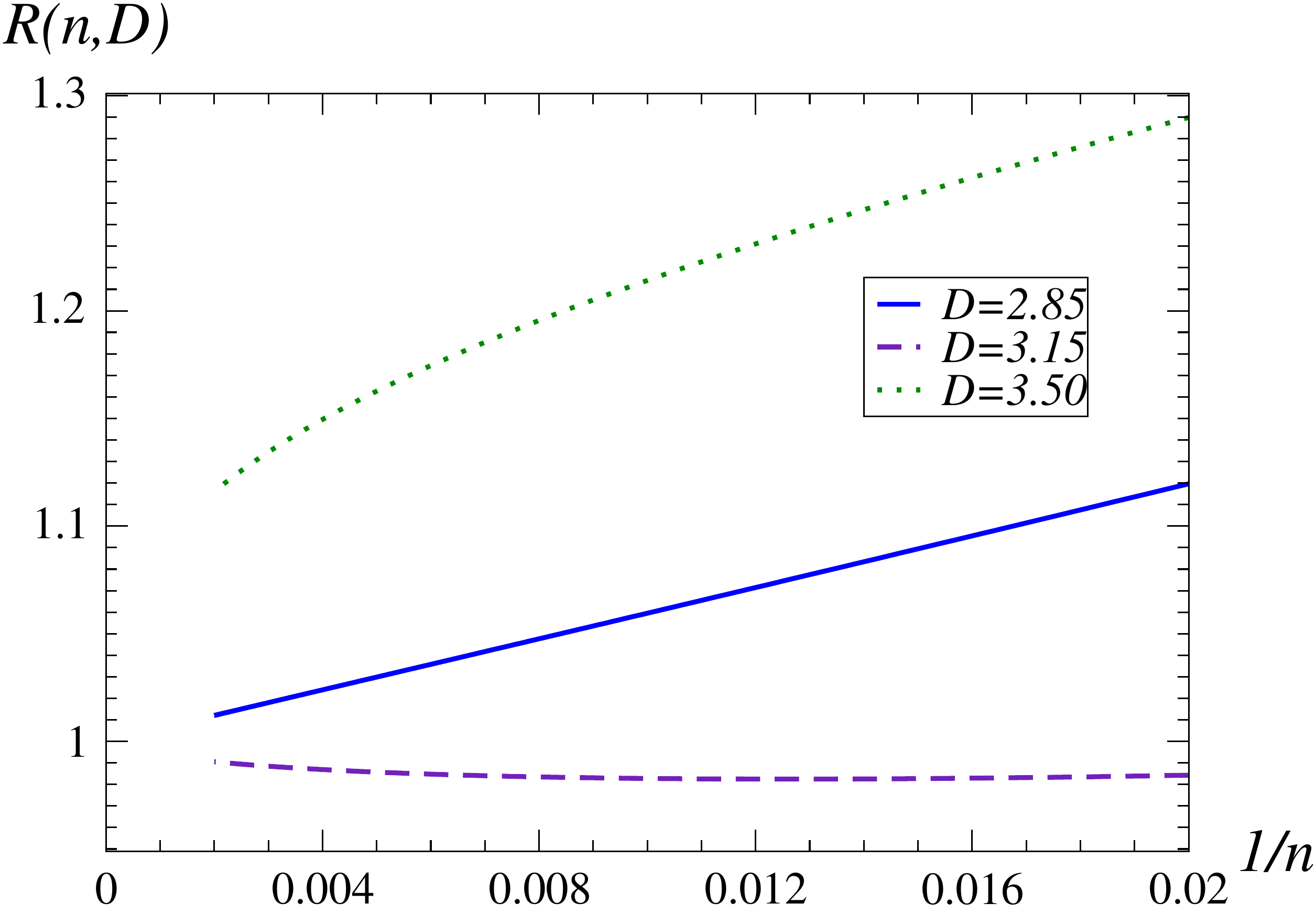}
  \caption{\label{BEcontribution}
 The ratios $R(n,d)= A_n(d)/A_n^{\text{BE}}(d)$
  as function of $n^{-1}$, plotted for $n=50,\ldots,500$.
  The $A_n(d)$'s are calculated numerically (using 750
  decimals accuracy), and divided by the {\em leading order\/}
  contribution (\ref{BoseEinsteinContribution}) to $A_n^{\text{BE}}(d)$.
  The first correction to this ratio is of order $n^{-1}$ when
  $\delta>1$ (i.e.\ for $d<3$), and of order $n^{1-\delta}$ when
  $\delta<1$. It is straightforward to calculate such corrections,
  but they are best read by computers.}
\end{figure}

We have checked the accuracy of~(\ref{BoseEinsteinContribution}) for a few non-special dimensions, see figure~\ref{BEcontribution}.
%
%
The higher order corrections have the form of a double series in powers of $n^{-1}$ and $n^{1-\delta}$. Since
$\delta\to1^+$ as $d\to4^-$ the convergence towards (\ref{BoseEinsteinContribution}) becomes slow near $d=4$,
as demonstrated by the $d=3.5$ case in Fig.~\ref{BEcontribution}.

It is now easy to understand why $A_n(d)=0$ for $d\approx d_m$. The contribution from (\ref{BoseEinsteinContribution}) 
vanishes like $(d-d_m)\, \rho_{\text{BE}}^{-n}$. It can be cancelled by the contribution from the singularities at
$\rho_{\pm}$. The latter behaves like $\vert \rho_{\pm} \vert^{-n}\,\cos(n\omega+\phi)$. Hence (ignoring algebraic 
prefactors) there will be a zero when
\begin{equation}
  d-d_m \sim \left(\frac{\rho_{\text{BE}}}{\vert\rho_{\pm} \vert}\right)^n\,\cos(n\omega+\phi).
\end{equation}

\section{Exploring dimensionality\label{ExploringDimension}}

%
\begin{figure}
  \includegraphics[]{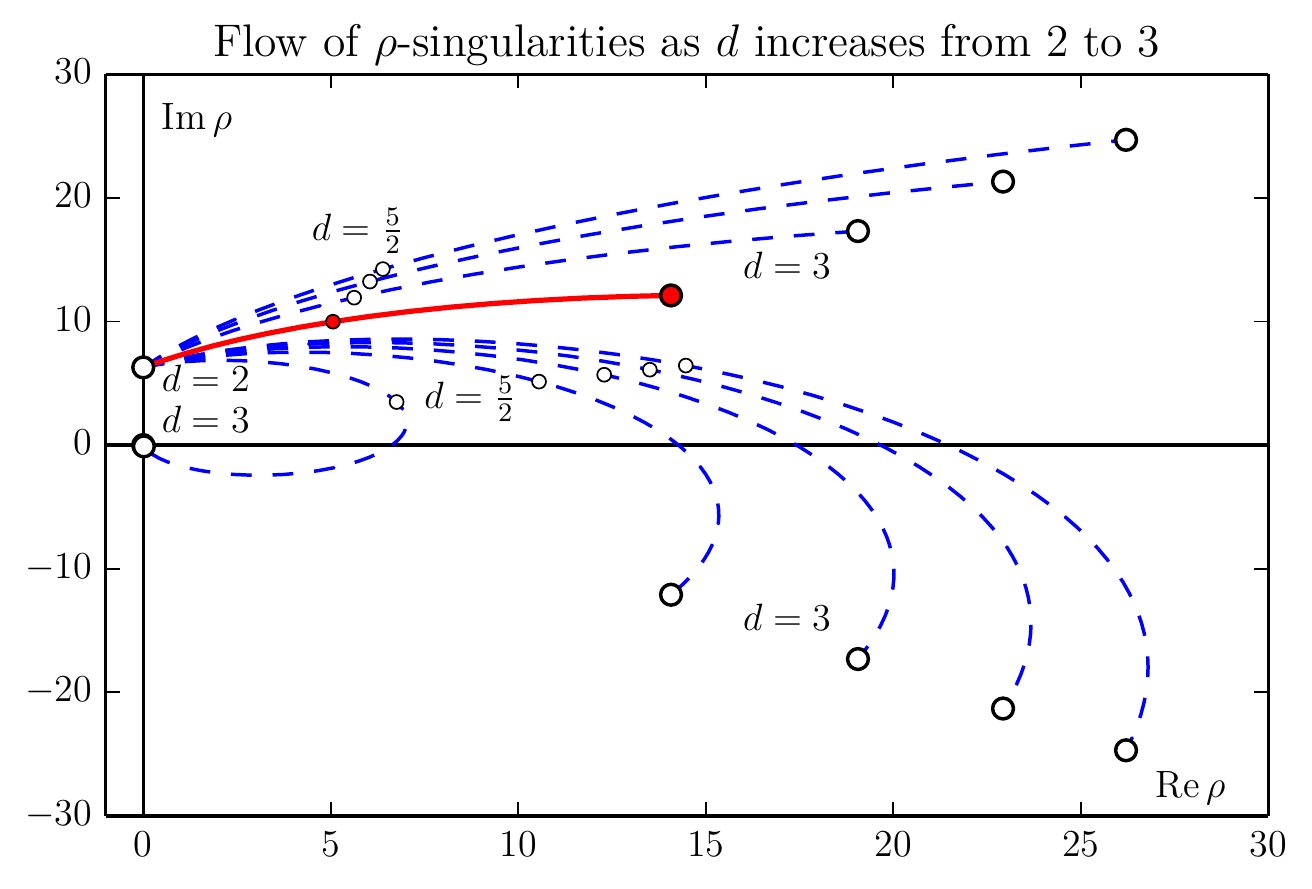}
  \caption{\label{path2to3} 
   This figure illustrates how a few (out of an
  infinite number) of singularities flow away from $\rho=2\pi\text{i}$
  as $d$ is increased from 2 to 3. There is a similar flow related by
  complex conjugation and $k_0=1 \to k_0\to -1$. For exactly $d=3$ the 
  latter distinction vanishes: The generically logarithmic
  singularities at $\mu=2\pi\text{i}m$ become square root singularities at exactly
  $d=3$. As a consequence a translation symmetry
  under $k\to k+2$ emerges. For this reason the apparent singularity at
  $\rho=0$ for $d=3$ is probabably annihilated by its complex
  conjugate, and thus not present at $d=3$.
  }
\end{figure}
%

For general dimension $\rho_{+}$ (and similar singularities) is determined by 
equation~(\ref{Second_Singularity}). Since we know a complex conjugate pair of
solutions for $d=3$, defined by $(\mu_+, k_0=1)$ and $(\mu_-=\mu_+^*,k_0=-1)$, 
it is straightforward to explore how they move with changing dimension, by changing 
$d$ in small steps and changing $\bm{k}$ when branch cuts are crossed. A map of this 
exploration (in the $z$-plane) is shown in Fig.~\ref{FindingRhoPlus}, and described 
in the last part of its caption. As $d=2$ the singularity approaches $z=0$, which in 
this case corresponds to $\rho=2\pi\text{i}$. The latter can be seen in Fig.~\ref{ChangingDimension}, 
where a map of the same curve is shown in the $\rho$-plane.

Actually, when approaching $d=2$ one encounters a puzzle. The singularity $\rho_+(d)$ is generally 
of square root type, with two Riemann sheets attached locally. However, for $d=2$ it is logarithmic 
with infinitely many attached Riemann sheets. How can a square root singularity suddenly turn into a
logarithmic one? The answer is that this is impossible for a single square root, but it may be possible 
if we have infinitely many of them. Hence, one must conclude that there are infinitely many $\rho$-plane 
singularities approaching $2\pi\text{i}n$ when $d\to2$. They are actually easy to locate by direct
analysis of equation~(\ref{Second_Singularity}). They correspond to $\text{Re}\,\mu \sim \log\vert d-2 \vert$,
and values of $\text{Im}\,\mu$ which change in steps of approximately $2\pi$.

%
\begin{figure}
  \begin{center}
    \includegraphics[]{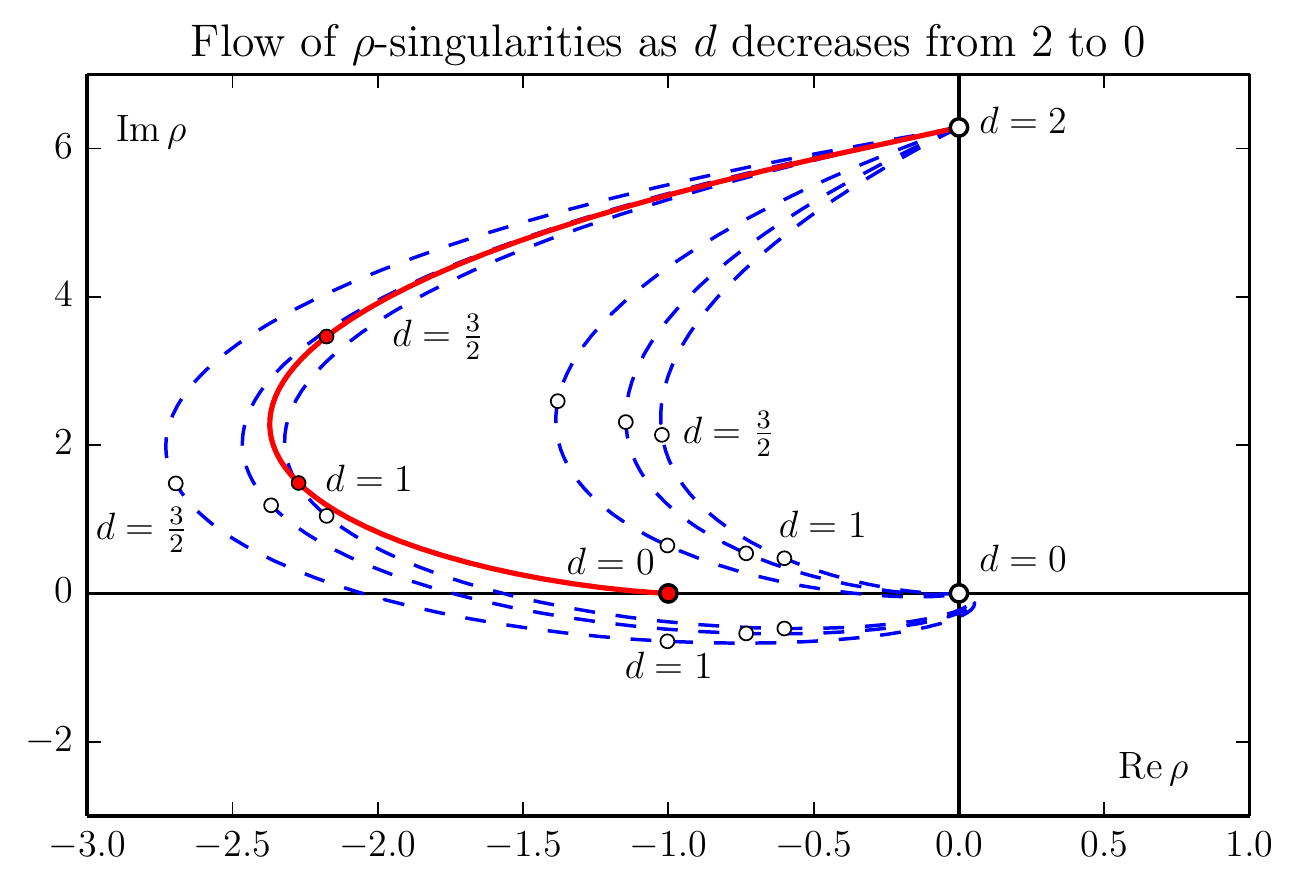}
\end{center}
\caption{\label{path2to0}
This figure illustrates how a few (out of an infinite number) of singularities flow away from $\rho=2\pi\text{i}$
as $d$ is decreased from 2 to 0. There is a similar flow related by
complex conjugation.}
\end{figure}
%

A simple way to generate them numerically is to follow the solution as $d$ encircles $d=2$ in 
the plane of complex dimensions, $d=2+\delta\,\text{e}^{\text{i}\phi}$ with some small 
positive $\delta$. Note that equation~(\ref{Second_Singularity}) has only pole singularities
as function of $d$, hence each circle leads back to the same equation. However, each change 
of $\phi$ by $2\pi$ leads to a new solution. Having generated many new singularities
this way one may again follow their paths back to $d=3$. This is shown in Fig.~\ref{path2to3}.

Similarly, we may change $\phi$ to $\phi+(2n+1)\pi$ for various $n$, and see how the singularities 
flow when we decrease $d$ from $2-\delta$ towards $d=0$. This is illustrated in Fig.\ref{path2to0}. 
To our initial surprise, most  singularities (probably infinitely many --- all but one) flow towards $\rho=0$.
On the other hand, we know from the explicitly found equation of state at $d=0$ that there is no 
singularity at $\rho=0$ on any Riemann sheet. Hence, it must be that they annihilate 
at $d=0$. By direct analysis of the equation it is possible to see that there must be 
an infinite number of solutions to (\ref{Second_Singularity}) approaching
$\rho=0$ as $d\to 0$. The analysis is similar to the one for $d=2$. We have not investigated 
the ``annihilation process'' in detail. Another surprise is that only one square root singularity 
flows toward the logarithmic one at $\rho=-1$ for $d=0$ (there is one more related by complex 
conjugation). The infinity of additional singularities needed to build a logarithmic one must 
flow from the other $d=2$ singular points $2\pi\text{i}m$.

\section{Summary and Conclusions\label{Conclusions}}

In this paper, we have investigated the analytical structure of the virial expansion and equation
of state of ideal quantum gases in arbitrary (real-valued) dimensions, with an emphasis on locating
the singularities that determine the radius of convergence of the expansion. These simple systems
have a surprisingly rich complex analytical structure.  

When investigating the behavior near $d=3$, one finds that {\em we live in a very special dimension\/} 
from the point of view of the virial expansion. Namely, for $d$ close to $3$ the equality $\rho_c(d)=\rho_{\text{BC}}(d)$ 
holds for every $d\ne3$, but {\em not\/} for exactly $d=3$. The virial coefficients $A_n(d)$ have zeros
which approach $d=3$ very rapidly as $n$ increases. This leads to a sudden increase of the
convergence radius at exactly 3 dimensions, paradoxically extending the radius of convergence far beyond the
critical density for Bose-Einstein  condensation (in the bosonic case). Enlarging the investigations to a wider 
$d$-range, one finds the  same behavior at a few other special dimensions $d_m$. There are six such cases in all,
$d_m=2+{2}/({m+1})$ for $m=1, 2 \ldots 6$. We have revealed a rich analytical structure in the virial expansion,
when suitably extended to the complex plane, that explains this behavior. 

When searching for the complex conjugate pairs of singularities which determine $\rho_c(d_m)$,
we find that they are determined by the simple equation~(\ref{SecondSingularity}) in $d=3$
dimensions, and more generally by equation~(\ref{Second_Singularity}).

While we have focused exclusively on noninteracting quantum gases in this paper, we expect that some
of the results may be extended to interacting cases. Interacting systems where mean-field theories are 
applicable (that is, above some lower critical dimension) are essentially effective single particle problems 
and as such may fall within the class of problems we have considered here, provided that the resulting mean-field 
theory features long-lived excitations above the ground state condensate which are either fermionic or bosonic. 
Mean field theories go beyond any order in perturbation theory and often capture interesting physics of strong-coupling 
fixed points. They tend to be exact as the dimension  approaches an upper critical dimension. In this context, we note 
here that the effective dimension $d$ is twice the dimension of physical space in the relativistic cases. Low-energy 
excitations of quantum spin systems defined on fractal lattices may also be described by dilute Bose gases with noninteger 
$d$.    

A.S. was supported by the Research Council of Norway, through Grants 205591/V20 and 216700/F20.

\end{document}